\documentstyle[psfig]{mn}
\begin{document}
\def\simlt{\mathrel{\rlap{\lower 3pt\hbox{$\sim$}}
        \raise 2.0pt\hbox{$<$}}}
\def\simgt{\mathrel{\rlap{\lower 3pt\hbox{$\sim$}}
        \raise 2.0pt\hbox{$>$}}}

\title[Evidence for anisotropy in the distribution of short-lived gamma-ray 
bursts]
{Evidence for anisotropy in the distribution of short-lived gamma-ray bursts}
\author[M. Magliocchetti, G. Ghirlanda  \& A. Celotti]
{M. Magliocchetti$^{1}$, G. Ghirlanda$^{2}$, A. Celotti$^1$ \\
$^1$SISSA, Via Beirut 4, I-34014, Trieste, Italy\\ 
$^2$Istituto di Fisica Cosmica - CNR, Via Bassini, 5, I-20133, Milano, Italy \\}

\maketitle
\begin{abstract}
Measurements of the two-point angular correlation function $w(\theta)$
for 407 short gamma-ray bursts collected in the Current BATSE
Catalogue reveal a $\sim$2$\sigma$ deviation from isotropy on angular
scales $\theta\sim 2-4$ degrees. Such an anisotropy is not observed in
the distribution of long gamma-ray bursts and hints to the presence of
repeated bursts for up to $\sim 13\%$ of the sources under
exam. However, the available data cannot exclude the signal as due to
the presence of large-scale structure.  Under this assumption, the
amplitude of the observed $w(\theta)$ is compatible with those derived
for different populations of galaxies up to redshifts $\simeq 0.5$,
result that suggests short gamma-ray bursts to be relatively local
sources.
\end{abstract}

\section{Introduction}

The Burst And Transient Source Experiment (BATSE) on board the Compton
Gamma Ray Observatory collected more than 2000 Gamma-Ray Bursts (GRBs)
in its 9-year activity. The analysis of such a large sample has deeply
characterized the temporal and spectral properties of GRBs (Fishman \&
Meegan 1995). Despite the multitude and complexity of the light curves
observed, the burst duration, typically defined by $T_{90}$ as the
time interval during which 90\% of the fluence is accumulated
(Kouvelitou et al. 1993), indicates the existence of two main classes:
short ($T_{90}<2$ s) and long ($T_{90}<2$ s) bursts. This distinction
is also reinforced by the associated spectral shape since short bursts
are typically harder than long ones (Fishman \& Meegan 1995;
Mitrofanov et al. 1998). At present, it is not known if this dichotomy
is produced by different progenitors or by different physical
mechanisms originating from the same object or, again, from different
observing conditions.  Nonetheless, it has been suggested (MacFadyen
\& Woosley 1999) that long bursts might be produced by the core
collapse of a single massive stars (the Collapsar model -- Woosley
1993; Paczynski 1998) whereas short bursts might be more efficiently
generated by double compact object mergers (Goodman 1997; Narayan,
Piran \& Kumar 2001) or double neutron star heating (Salmonson \&
Wilson 2002). A distinctive feature of these two scenarios is the
expected GRB distance: compact object mergers should have a lower
median redshift than Collapsars (Fryer, Woosley \& Hartmann 1999)
which would be tracers of the high redshift universe (Lamb 2001).

A cosmological origin of long bursts was first suggested by the
isotropic sky distribution observed by BATSE (Briggs 1993; Meegan et
al. 1992; Paczynski 1995) but the definitive proof came with the first
redshift measurement for GRB 970508 (Metzger et al. 1997) at
$z=0.835$. At present, the distribution of known redshifts extends up
to z=4.5 with a mean $\langle z \rangle\sim 1.27$ (Andersen et
al. 2000; Djorgovski et al. 2001). However very little is presently
known about short GRBs because no radio, optical or X-ray counterpart
has been associated with any burst lasting less than 2 s (Hurley et
al. 2002), although the possible existence of short burst afterglows
has been suggested (Lazzati, Ramirez-Ruiz \& Ghisellini 2001). Thus,
the distance scale of short GRBs is still an open issue and its
solution would have important implications for the understanding of
their progenitors.

The aim of this Letter is to investigate the nature of short GRBs by
means of statistical techniques such as the angular two-point
correlation function, since different hypotheses are expected to leave
different imprints on their distribution on the sky. In particular we
will concentrate on the two distinct possibilities of burst repetition
and large-scale structure.\\ In order to do so, we have considered
sources from the public BATSE on line
catalogue\footnote{http://gammaray.msfc.nasa.gov/batse/grb/catalog/current/}
which contains 2702 bursts from GRB 910421 to GRB 000526. For our
analysis, a key ingredient is the estimate of the duration $T_{90}$,
which has been derived only for 2041 events of this catalogue. Also,
the flux and fluence are only available for 2135 events. The maximum
number of bursts that we could extract with measured durations and
fluences is 2035, of which 497 are short events with $T_{90}<2$ s.

Our work is presented as follows: in Section 2 we briefly recall the
results of previous works examining evidence for GRB anisotropies,
while in Section 3 we introduce the method and show the results on the
angular correlation function.  Section 4 discusses some possible
models for the observed distribution of short GRBs and summarizes our
conclusions.

\section{Evidence of GRB anisotropies}

First studies on the angular distribution of GRBs by Hartmann \&
Blumenthal (1989) including 80 GRBs detected by the IPN and 160 GRBs
collected by KONUS, showed a possible deviation from isotropy on
scales $\sim 5^{\circ}$. A more refined analysis (Quashnock \& Lamb
1993) based on the first BATSE (1B) catalogue (260 bursts) confirmed
this evidence but with a cut on the burst positional errors
$\Delta<10^{\circ}$ (but see cautions in Narayan \& Piran 1993;
Brainerd 1996).

The analysis of joint spatial and temporal clustering of the bursts in
the 1B catalog (Wang \& Lingenfelter 1995) showed an excess of bursts
with angular separation lower than $4^{\circ}$ within 4 days with a
small probability ($1 \times 10^{-5}$) of random occurrence. However,
less significant results are claimed by Petrosian \& Efron (1995).
In particular, the need of larger samples for investigation of the
possible burst repetition by means of the two--point correlation
function is discussed by Brainerd (1996).

Evidence of possible burst clustering was recently claimed 
for both the populations of short and intermediate GRBs from the analysis of 
larger samples (Balazs et al. 1998; Meszaros et al. 2000). The possible 
different nature of the short and long events, in fact suggests the separate 
analysis and comparison of their clustering properties.


\section{The Method}

One of the possible statistical tests for the detection of anisotropy
is provided by the two-point angular correlation function $w(\theta)$,
which gives the excess probability, with respect to a random Poisson
distribution, of finding two sources in the solid angles
$\delta\Omega_1$ $\delta\Omega_2$ separated by an angle $\theta.$ We
applied this technique to the distinct samples of long and short GRBs
in order to investigate for possible deviations from isotropy in their
spatial distribution.

\begin{figure}
\vspace{8cm}  
\includegraphics{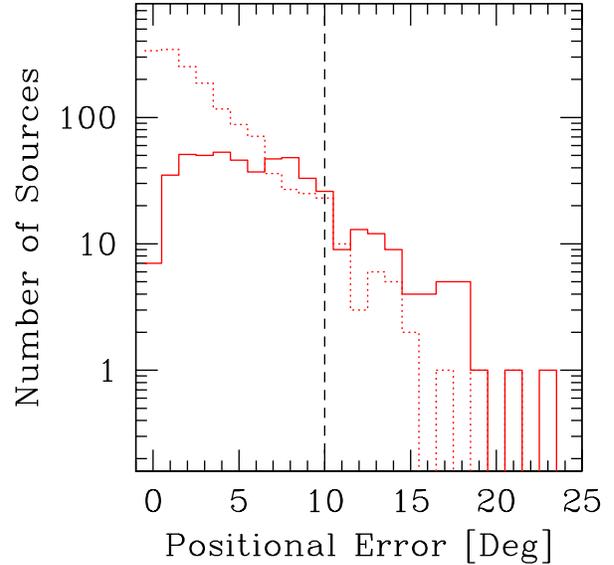}
\caption{Distribution of positional errors for objects included in the
Current BATSE Catalogue. The dotted line is for the population of
long-lived GRBs, while the solid one is for short-lived bursts. The
vertical line indicates the upper limit for inclusion in the datasets
considered in this work (see text for details).}
\label{fig:hist_err}
\end{figure}

Given the large uncertainties affecting the determination of the
positions of GRBs, especially in the case of short-lived ones (see
Figure 1), we have decided to only consider those objects which have
angular coordinates with associated errors $\Delta \le 10^{\circ}$. The
choice for this value results as the best compromise between the
requirements of precision-position measurements and large-enough
datasets to allow for sensible statistical analysis.  By applying the
above cut, we end up respectively with $N_{\rm long}= 1486$ (out
of 1538) and $N_{\rm short}=407$ (out of 497) sources in the case of
long-lived and short-lived GRBs.

Random catalogues have subsequently been generated for each of the two
subsets, with positions of the objects uniformly distributed on the
celestial sphere. In order to minimize Poisson fluctuations due to the
limited number of sources in the original datasets, random catalogues
were required to include $10\times N_{\rm long}$ objects in the case
of long-lived GRBs and $20\times N_{\rm short}$ objects in the case of
short-lived GRBs.  We then counted the number of distinct data-data
pairs ($DD$), data-random pairs ($DR)$, and distinct random-random
pairs ($RR$) as a function of angular separation, and calculated the
two-point angular correlation function $w(\theta)$ using the estimator
(Hamilton 1993)
\begin{eqnarray}
w(\theta) = &\frac{4DD \ RR}{(DR)^2}  &-1
\label{eqn:wtheta_ests}
\end{eqnarray}
on the angular scales $1\le \theta \le 20$ degrees.

\begin{figure}
\vspace{9cm}  
\includegraphics{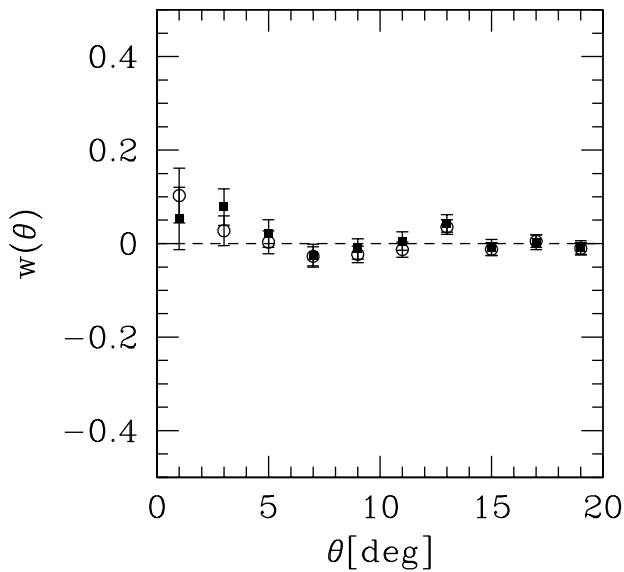}
\caption{Angular correlation function $w(\theta)$ of long-lived
gamma-ray bursts from the Current BATSE Catalogue. Empty dots are for
sources with positional uncertainties $\Delta\le 10^\circ$ (1486
objects), while filled squares are for sources with positional
uncertainties $\Delta\le 5^\circ$ (1239 objects).}
\label{fig:w_long}
\end{figure}

\begin{figure}
\vspace{9cm}  
\includegraphics{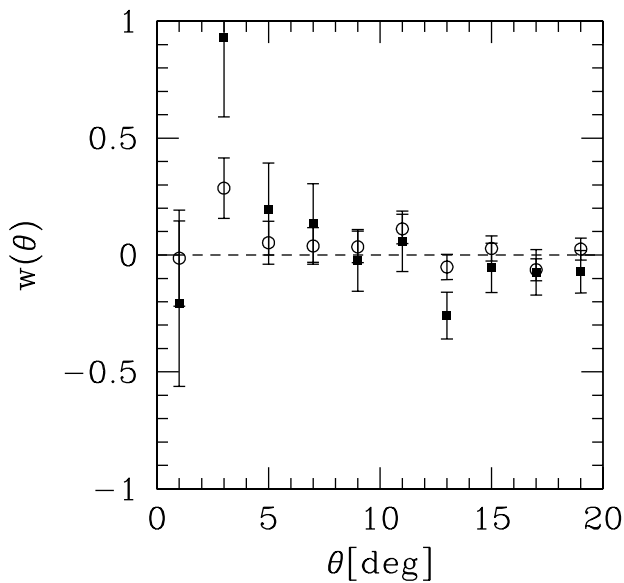}
\caption{Angular correlation function $w(\theta)$ of short-lived
gamma-ray bursts from the Current BATSE Catalogue. Empty dots are for
sources with positional uncertainties $\Delta\le 10^\circ$ (407
objects), while filled squares are for sources with positional
uncertainties $\Delta\le 5^\circ$ (196 objects).}
\label{fig:w_short}
\end{figure}

The open dots in Figs. 2 and 3 illustrate the results respectively for
the population of long-lived and short-lived GRBs; the error bars
indicate Poisson estimates for the points. We note that, while in the
first case values of $w(\theta)$ at all angular scales seem to be
consistent with an isotropic distribution, the angular correlation
function of short-lived GRBs shows a significant excess of power on
scales $\sim 2-4$ degrees.

In order to test the significance of this result, we have furtherly
examined two reduced datasets, obtained from the original one by only
considering those sources with positions known to better than
$\Delta\le 5^\circ$ (1239 long and 196 short GRBs). Solid squares in
both Figs. 2 and 3 illustrate the values of $w(\theta)$ obtained in
this case. Once again, measurements of the angular correlation
function for long-lived GRBs are entirely consistent with absence of
intrinsic non-randomness, in agreement with earlier results
(e.g. Meegan et al. 1995, 1996; Bennett \& Rhie 1996; Tegmark et
al. 1996) obtained for samples which mainly included this population
of gamma-ray emitters, and with more recent findings (e.g. Balazs et
al. 1998; Meszaros et al. 2000) concerning the population of
long-lived GRBs alone.
 
On the other hand, the distribution of short-lived, $\Delta\le
5^\circ$ GRBs -- even though with larger errors due to the small
number of objects considered in the analysis -- shows a departure from
isotropy on small angular scales even stronger than it was for the
previously-analysed $\Delta\le 10^\circ$ subsample. We can then
conclude that the distribution of short-lived gamma-ray bursts
exhibits a $\sim$2$\sigma$ deviation from isotropy on scales $\sim
2-4$ degrees.

\section{DISCUSSION}

Given the above result on the anisotropic distribution of short-lived
GRBs, it is worthwhile investigating (and possibly excluding) all the
systematics which could produce a non-zero signal in the angular
correlation function.\\ One of the possible causes for systematic
deviations from isotropy is related to the non-uniform sky-exposure of
BATSE (Fishman et al. 1994; Band 1996), which in principle could lead
to measurements of anisotropy even for randomly distributed
sources. We note however that such an effect would affect the observed
distribution of both long and short GRBs in a similar fashion, since
there is no {\it a priori} reason why different sub-classes should be
observed by the instrument in different ways. The fact that long-lived
sources exhibit a $w(\theta)$ which is highly consistent with the
hypothesis of isotropy at all the angular scales probed by our
analysis then tends to exclude responsibility from instrumental
effects on the observed anisotropy of short-lived GRBs (see also
Balazs et al. 1998; Meszaros et al. 2000).

Another possible cause for the observed anisotropic distribution can
be attributed to intrinsic Poissonian fluctuations of short-lived GRBs
due to their small number. Such an effect would however produce random
fluctuations of the measured correlation function at all angular
scales, while the data in our case show that an appreciable deviation
from zero values is only obtained on small distances; measurements of
$w(\theta)$ at all $\theta\simgt 5^\circ$ are remarkably consistent
with a null clustering signal. As a further test, we have re-binned
sources belonging to the sub-class of short-lived GRBs and evaluated
the two-point correlation function by means of equation
(\ref{eqn:wtheta_ests}) in the angular range $1\le \theta \le 10$
degrees.\\ Results from this analysis are presented in Figure 4: once
again a marked deviation from isotropy is only detected on scales
$\theta\simeq 2-4$ degrees, in full agreement with the findings of the
previous Section.

\begin{figure}
\vspace{8cm}  
\includegraphics{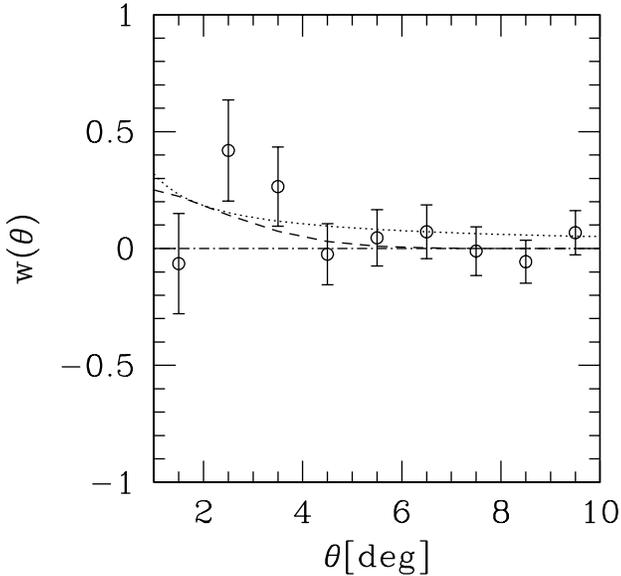}
\caption{Angular correlation function $w(\theta)$ for short-lived
gamma-ray bursts with positional uncertainties $\Delta \le
10^\circ$. The dashed line shows the best fit to the data obtained
under the hypothesis of burst-repetition, while the dotted line
illustrates the best case for large-scale clustering (see text for
details).}
\label{fig:w_short_10}
\end{figure}

Having excluded the effects of systematics on the detected anisotropy,
we are then left with intrinsic non-randomness in the distribution of
short-lived GRBs. This intrinsic anisotropic distribution would arise
if e.g. some of the sources were repeating bursts, as indeed expected
in several GRB models (Quashnock \& Lamb 1993; Wang \& Lingenfelter
1995; Meegan et al. 1995; Bennett \& Rie 1996; Tegmark et al.  1996).

If we define $f$ as the fraction of observed bursts that can be
labeled as repeaters and $\nu$ as the average number of observed
events per repeating source ($\nu\ge 2$) then, for positional errors
drawn from a Gaussian distribution with standard deviation
$\Sigma=\Delta/(2\sqrt{2 {\rm ln}2}) < < 1$ ($\Delta\simeq 5^\circ$,
see Figure 1), the correlation signal arising from the repeating-burst
assumption can be written as (see Meegan et al. 1995)
\begin{equation}
w(\theta)=\frac{f\:(\nu-1)}{(N_{\rm short}-1)}\;\left[\frac{2}{\Sigma^2}
{\rm exp}\left(-\frac{\theta^2}{2\Sigma^2}\right)-1\right].
\label{eq:w}
\end{equation} 
Equation (\ref{eq:w}) was used to fit the data presented in Figure 4
for different values of the product $f\: (\nu-1)$. It turns out that
the best-fit value is given by $f\: (\nu-1)=0.07^{+0.06}_{-0.04}$,
result which rejects the hypothesis of non-repetitivity at the $\sim
2\sigma$ confidence level.\\ By then fixing the average number of
repeating events, we obtain the following constraints on the fraction
of sources expected to repeat:
\begin{eqnarray}
\nu=2 \rightarrow f=0.07^{+0.06}_{-0.04}\;\;\;\nonumber\\
\nu=3 \rightarrow f=0.035^{+0.035}_{-0.02}\nonumber\\
\nu=4 \rightarrow f=0.023^{+0.02}_{-0.013}\nonumber
\end{eqnarray}  
In other words, the data indicate that up to 13 per cent of short
GRBs may have burst on average twice during the years corresponding to
the lifetime of BATSE. This fraction obviously decreases for an
increasing number of repetitions, going down to a more
negligible $\sim 2$ per cent in the case of $\nu=4$.

The above explanation is however not the only one capable to account
for the observed anisotropy on a few degrees scale. Another
alternative which needs to be taken into account is the presence of
large-scale structure, arising under the hypothesis that short GRBs
trace the distribution of the galaxies in which they reside (note that
here we are explicitly assuming GRBs to be extragalactic sources).  In
this framework, the angular correlation function can be modeled (to a
first approximation) as the power-law $w(\theta)=A\;
\theta^{(1-\gamma)}$ (see e.g. Peebles 1980).

Fits to the data presented in Figure 4 give values for the amplitude
of the correlation function $A\simeq 0.3\pm 0.2$, independent (within
the errors) of the different values for the slope
$\gamma=1.6,2.0,1.8$, chosen so to mimic the clustering properties
respectively of star-forming and early-type galaxies, and of a mixture
of both. The case for $\gamma=1.8$ is illustrated in Figure 4 by the
dotted line. Despite the large errors associated to its determination,
the allowed range of values for $A$ is in good agreement with results
e.g.  from the APM Survey (Loveday et al. 1995) based on samples of
$z\simlt 0.2$ galaxies divided by morphological type. On the other
hand, clustering analyses performed at higher redshifts find
amplitudes which are much smaller than the values indicated by the GRB
data ($A\simlt 10^{-2}$ for $z\simeq 1-1.5$ EROs, Roche et al. 2002;
$A\simeq 10^{-3}$ for $z\sim 3$ Lyman Break galaxies, Porciani \&
Giavalisco 2002; $A\simeq 5 \cdot 10^{-3}$ for $z\sim 2.5-3.5$
galaxies found in the Hubble Deep Field, Magliocchetti \& Maddox
1999).  If we then use large-scale structure to explain the clustering
measurements presented in this work, we can conclude that -- if
confirmed -- they imply short-lived GRBs to be hosted by relatively
(possibly $z\simlt 0.5$) local galaxies, in agreement with the median
redshift 0.5--0.8 predicted by neutron star/neutron star or black
hole/neutron star merger models (Fryer et al. 1999). In fact (see
e.g. MacFadyen \& Woosley 1999), the alternative to produce short
bursts in the context of the Collapsar model would meet some
difficulties at $z\simlt 0.8$ since the intense mass loss from the
progenitor - due to a higher metallicity with respect to that of stars
at higher redshifts - could prevent the collapse of the helium core.

Obviously, one would like to discern between an anisotropic signal
caused by burst repetition and one due to large-scale
structure. Unfortunately, the still small number of detected sources
does not allow to probe angular scales $\theta\simlt 1$ degrees where
these two effects would greatly differentiate from one another. 
The solution to this problem is however within reach, since the goal
of collecting hundreds of GRBs per year will be soon achieved by SWIFT
(Burrows et al. 2002), which will also allow for precise position
measurements in order to perform optical to X-ray spectrophotometry,
with the aim to determine the origin of GRBs and their afterglows and
to use bursts to investigate the early Universe. Moreover, the likely
detection of several short bursts (which was severely limited in the
BATSE experiment due to its lower sensitivity to extremely fast
transients) by INTEGRAL and their location with sufficient accuracy
through the IPN system will lead to a better understanding of the
still mysterious population of short bursts.

\end{document}